\documentclass[twocolumn,notitlepage,prc,showpacs,preprintnumbers,nofootinbib,superscriptaddress,floatfix,balancelastpage]{revtex4}
\pdfoutput=1
\usepackage{mathrsfs}
\usepackage{amssymb}
\usepackage{amsmath}
\usepackage{graphicx}
\usepackage{hyperref}
\usepackage{amsfonts,amsbsy}
\usepackage{animate}
\usepackage{color}
\usepackage{mciteplus}

\newcommand{\rt}{{\mathbf{r}_T}}
\newcommand{\xt}{{\mathbf{x}_T}}
\newcommand{\ut}{{\mathbf{u}_T}}
\newcommand{\xbt}{{\mathbf{{\bar x}}_T}}
\newcommand{\ybt}{{\mathbf{{\bar y}}_T}}
\newcommand{\zbt}{{\mathbf{{\bar z}}_T}}
\newcommand{\vt}{{\mathbf{v}_T}}
\newcommand{\bt}{{\mathbf{b}_T}}
\newcommand{\yt}{{\mathbf{y}_T}}
\newcommand{\zt}{{\mathbf{z}_T}}

\newcommand{\kt}{{\mathbf{k}_T}}

\newcommand{\ud}{\, \mathrm{d}}

\newcommand{\tr}{\, \mathrm{Tr} \, }

\newcommand{\nc}{{N_\mathrm{c}}}

\newcommand{\cf}{C_\mathrm{F}}

\newcommand{\nr}[1]{(\ref{#1})}

\newcommand{\qs}{Q_\mathrm{s}}

\newcommand{\as}{\alpha_{\mathrm{s}}}

\newcommand{\fig}{Fig.~}

\newcommand{\eq}{Eq.~}

\newcommand{\eqs}{Eqs.~}

\begin{document}

\author{A. Dumitru}
\affiliation{RIKEN BNL Research Center, Brookhaven National Laboratory,
Upton, NY-11973, USA
}
\affiliation{Department of Natural Sciences, Baruch College, CUNY, 17 Lexington Avenue, New York, NY 10010, USA
}
\affiliation{The Graduate School and University Center, CUNY, 365 Fifth Avenue, New York, NY 10016, USA
}
\author{J. Jalilian-Marian}
\affiliation{Department of Natural Sciences, Baruch College, CUNY, 17 Lexington Avenue, New York, NY 10010, USA
}
\affiliation{The Graduate School and University Center, CUNY, 365 Fifth Avenue, New York, NY 10016, USA
}
\author{T. Lappi}
\affiliation{
Department of Physics, %
 P.O. Box 35, 40014 University of Jyv\"askyl\"a, Finland}
\affiliation{
Helsinki Institute of Physics, P.O. Box 64, 00014 University of Helsinki,
Finland}
\author{B. Schenke}
\affiliation{Physics Department,   Brookhaven National Laboratory,
  Upton, NY-11973, USA
}
\author{R. Venugopalan}
\affiliation{Physics Department,   Brookhaven National Laboratory,
  Upton, NY-11973, USA
}

\title{
Renormalization group evolution of multi-gluon correlators in high energy QCD
}

\preprint{RBRC-914}

\begin{abstract}
Many-body QCD in leading high energy Regge asymptotics is described by the 
Balitsky-JIMWLK  hierarchy of renormalization group  equations for the $x$
 evolution of multi-point Wilson line correlators. These correlators are 
universal and ubiquitous in final states in deeply inelastic scattering and 
hadronic collisions.  For instance, recently measured di-hadron correlations
 at forward rapidity in deuteron-gold collisions at the Relativistic Heavy
 Ion Collider (RHIC) are sensitive to four and six point correlators of 
Wilson lines in the small $x$ color fields of the dense nuclear target. 
We evaluate these correlators numerically by solving the functional
 Langevin equation that describes the  Balitsky-JIMWLK hierarchy. 
We compare the results to mean-field Gaussian and large $\nc$ 
approximations used in previous phenomenological studies. We 
comment on the implications of our results for quantitative studies 
of multi-gluon final states in high energy QCD.

\end{abstract}

\maketitle

\section{Introduction}

QCD in high energy Regge asymptotics can be described as a dense many-body 
system of ``wee'' gluons and sea quarks.  In the infinite momentum frame, 
gluons with transverse momenta $k_\perp \lesssim \qs$ saturate phase space
 maximally, where $\qs (x)$ is a dynamical saturation scale~\cite{Gribov:1984tu,*Mueller:1985wy} 
that grows with decreasing fractions $x$ of the longitudinal momentum 
of the hadron carried by the gluons. The properties of saturated gluons 
are described by the Color Glass Condensate (CGC) effective 
theory~\cite{Iancu:2003xm,*Gelis:2010nm,Weigert:2005us},
 where the degrees of freedom are static
 color sources in the hadron at large $x$, coupled to the dynamical 
wee gluon fields at small $x$. Renormalization group equations, 
derived from requiring that observables be independent of the
 separation in $x$ between sources and fields, lead to an 
infinite hierarchy of evolution equations in $x$, for n-point
 Wilson line correlators averaged over dense color fields in 
the hadron. Given appropriate initial conditions at large $x$,
 solutions of this Balitsky-JIMWLK hierarchy~\cite{Balitsky:1995ub,%
Jalilian-Marian:1997jx,*Jalilian-Marian:1997gr,%
*Iancu:2000hn,*Ferreiro:2001qy,*Mueller:2001uk} 
allow one to compute a wide range of multi-particle final 
states in deeply inelastic scattering (DIS) and hadronic collisions. 

A prominent example is provided by inclusive DIS structure functions 
$F_{2}$ and $F_{L}$, which are proportional to the forward 
scattering amplitude of a $q\bar{q}$  ``dipole'' on a nucleus. 
The forward dipole amplitude (dipole cross section) can be expressed as 
\begin{multline}
\sigma_\mathrm{dip.}(x, \rt) = 2\int \ud^2 \bt 
\\
\times
\bigg< 1 
- \frac{1}{\nc}
\tr  V\left(\bt + \frac{\rt}{2}\right)
V^\dagger\left(\bt - \frac{\rt}{2}\right) \bigg> \, ,
\label{eq:dipole}
\end{multline}
where $\rt = \xt -\yt$ is the transverse size of the dipole, 
$\bt = (\xt+\yt)/2$ is the impact parameter relative to the hadron, 
and the rapidity $Y = \ln(x_0/x)$, where $x_0$ is the initial scale for 
small x evolution. The dipole amplitude is the expectation value 
$D \equiv \langle \hat{D}\rangle$
of the  dipole operator 
\begin{equation}
\hat{D}(\xt-\yt) \equiv \frac{1}{\nc} \tr V(\xt) V^\dagger(\yt)
\end{equation}
averaged over the color sources of the target evaluated at the rapidity $Y$.
This average obeys the Balitsky-JIMWLK equation that relates
its energy dependence to the expectation value of a four-point operator:
\begin{multline} \label{eq:hierarchy}
 {\ud\over \ud Y} D (\xt - \yt) =  
{\nc\, \as \over 2\pi^2} 
\int_\zt
{(\xt - \yt)^2 \over (\xt - \zt)^2 (\zt - \yt)^2}\;
\\
\times 
\left\langle \hat{D} (\xt - \zt)\, \hat{D} (\zt - \yt) - \hat{D} 
(\xt - \yt)\right\rangle.
\end{multline}
In the large $\nc$ approximation 
the expectation value of $\hat{D}^2$
factorizes and the equation becomes a closed one. This is known as the 
 Balitsky-Kovchegov (BK) equation~\cite{Balitsky:1995ub,Kovchegov:1999yj}:
\begin{multline}
 {\ud\over \ud Y} D (\xt - \yt) =  
{\nc\, \as \over 2\pi^2} 
\int_\zt
{(\xt - \yt)^2 \over (\xt - \zt)^2 (\zt - \yt)^2}\;
\\
\times 
\left[D (\xt - \zt)\, D (\zt - \yt) - D (\xt - \yt)\right].
\label{eq:2pt}
\end{multline}

In addition to \eq\nr{eq:dipole}, this dipole correlator appears
in a number of final states in both DIS and hadronic scattering; the
BK equation for its energy evolution is widely used in
phenomenological applications. The mean field approximation 
$\langle\hat{D}^2\rangle \approx\langle\hat{D}\rangle^2$
has been checked by numerical
solutions of the JIMWLK equations, and it is seen that it is satisfied
to a very good approximation (much better than the $1/\nc^2$ one might
expect)~\cite{Rummukainen:2003ns,Kovchegov:2008mk}. Unless noted
otherwise, throughout the paper we define the saturation scale
$\qs(Y)$ from the expectation value of the dipole operator as
$D\left(r=\sqrt{2}/\qs\right) = e^{-1/2}$.

For less inclusive observables, new universal degrees of freedom
beyond dipoles are encountered. Examples include small-$x$ di-jet
production in e+A DIS~\cite{Dominguez:2011wm}, quark-antiquark pair
production in hadronic collisions~\cite{Blaizot:2004wv} and near-side
long-range rapidity correlations~\cite{Dumitru:2010mv}. Here we focus
on $n$-point functions which appear in forward
di-hadron production in light on heavy hadron collisions,
$p+A\longrightarrow h_1\;h_2\;X$, a process that has been studied
recently in deuteron-gold collisions at RHIC. When both hadrons are
produced at forward rapidities in the proton/deuteron fragmentation
region, the dominant underlying QCD process is the scattering of a
large $x_1$ valence quark from the deuteron off small $x_2$ partons in
the nuclear target, with the emission of a gluon from the valence
quark either before or after the collision. This cross-section is
expressed as~\cite{Marquet:2007vb,Dominguez:2011wm},
\begin{widetext}
\begin{multline}
{\ud \sigma^{qA\rightarrow qgX}\over \ud^3 k_1\;\ud^3 k_2} \propto {\as \nc\over 2} 
\int_{\xt,\xbt,\yt,\ybt}\hspace{-1.4cm} 
e^{-i\kt_1 \cdot (\xt-\xbt)}\;e^{-i \kt_2\cdot (\yt-\ybt)}\;
{\cal F}(\xbt-\ybt,\xt-\yt)
	  \Bigg\langle 
\hat{Q}(\yt,\ybt,\xbt,\xt)\; \hat{D}(\xt,\xbt) \\
- \hat{D}(\yt,\xt)\hat{D}(\xt,\zbt) 
		\hspace{-0.05cm}-\hspace{-0.05cm}
\hat{D}(\zt,\xbt) \hat{D}(\xbt,\ybt) 
		\hspace{-0.05cm}+\hspace{-0.05cm} 
\frac{C_F}{\nc} \hat{D}(\zt,\zbt)
		\hspace{-0.05cm} +\hspace{-0.05cm} 
\frac{1}{ \nc^2} \left(\hat{D}(\yt,\zbt) 
		\hspace{-0.05cm}+\hspace{-0.05cm} 
\hat{D}(\zt,\ybt)   
		\hspace{-0.05cm}-\hspace{-0.05cm}
\hat{D}(\yt,\ybt)\right) 
		\Bigg\rangle\,,
\label{eq:pA}
\end{multline}
\end{widetext}
with $\zt = z \xt +(1-z) \yt$ and likewise, $\zbt = z \xbt + (1-z)
\ybt$.
${\cal F}$ denotes the splitting function for producing a photon off a
quark, with the four co-ordinates denoting the transverse spatial
co-ordinates of the quark and gluon in the amplitude and conjugate
amplitude. The color field dynamics specific to gluon emission are absorbed in
the expectation value $\langle \rangle$, which contains a new quadrupole operator,
\begin{equation}
\hat{Q}(\xt,\yt,\ut,\vt) = {1\over \nc} \tr V(\xt) V^\dagger(\yt) 
V(\ut) V^\dagger(\vt).
\label{eq:quad}
\end{equation}
We denote the expectation value of the quadrupole operator
by $Q=\langle \hat{Q}\rangle$.
Unlike the $\langle \hat{D}^2\rangle $ in \eq\nr{eq:hierarchy}, 
it is not reducible to the product of
dipoles even in the large $\nc$ and large $A$ approximations and is a
novel universal correlator in high energy
QCD~\cite{JalilianMarian:2004da,Dominguez:2011wm}, interesting
both from theoretical and phenomenological perspectives.

In this paper, we determine the expectation values of relevant
Wilson line correlators for a SU(3) gauge group explicitly
numerically. It is known that the evolution of the expectation values of Wilson line
correlators can be expressed as a functional Fokker-Planck
equation~\cite{Weigert:2000gi}, which in turn can be re-expressed as a
functional Boltzmann-Langevin equation for the Wilson lines
themselves~\cite{Blaizot:2002xy},
\begin{equation}
\frac{\ud V(\xt)}{\ud Y} = V(\xt) (i t^a) \left\{ \sigma(\xt)^a+
\int_\zt \hspace{-0.25cm}
\varepsilon(\xt,\zt)^{ab}_i \; \xi(\zt)^b_i  
\right\}
\label{eq:Langevin}
\end{equation}
where 
\begin{equation}
 \varepsilon(\xt,\zt)^{ab}_i = \left({\as\over \pi}\right)^{1/2}\;
{(\xt-\zt)_i\over (\xt-\zt)^2}\;\left[1-U(\xt)^\dagger U(\zt)\right]^{ab} 
\label{eq:sqrt}
\end{equation}
is the ``square root'' of the  JIMWLK Hamiltonian. Here
$U$'s are adjoint Wilson lines, which are related to fundamental
Wilson lines through the identity, $U_{ab}(\xt)= 2 \tr ( t^a V^\dag(\xt) t^b
V (\xt))$.  The equation includes a term proportional to a 
Gaussian white noise $\xi$ satisfying $\langle \xi(\xt)^b_i\rangle
=0$ and $\langle \xi(\xt)^a_i \xi(\yt)^b_j\rangle = \delta^{ab}
\delta^{ij}\delta^{(2)}(\xt-\yt)$. Finally, there is also a 
drag term   
\begin{equation}
\sigma(\xt)^a = -i \frac{\as} {2\pi^2}
\int_{\zt} {1\over (\xt -\zt)^2} \tr \left[ T^a U(\xt)^\dagger U(\zt)\right],
\end{equation}
where $T^a$ is a generator of the adjoint representation.

\section{Numerical method}

As alluded to previously, numerical solutions of the
Boltzmann-Langevin hierarchy were obtained for fixed
coupling~\cite{Rummukainen:2003ns} and used to study the factorization
of dipole correlators. Apart from the running coupling,
we use the numerical method of ref.~\cite{Rummukainen:2003ns}
for solving the JIMWLK equation.
Several running coupling prescriptions have
been suggested in the
literature~\cite{Balitsky:2008zza,*Kovchegov:2006vj}; in this paper,
as in \cite{Lappi:2011ju}, we
will assume that the coupling constant runs as a function of the
``daughter'' dipole size $r = |\xt -\zt|$. The Landau pole is regulated 
by taking
\begin{equation}
  \as(r) = \frac{4\pi}{\beta \ln\left\{\left[\left(\frac{\mu_0^2}{\Lambda^2}\right)^{\frac{1}{c}}
      +\left(\frac{4}{r^2\Lambda^2}\right)^{\frac{1}{c}}\right]^{c}
\right\}}\,,
\end{equation}
with a parameter $c$ which regulates the sharpness of the cutoff. 
The scale $\Lambda$ in the coupling
is parametrically of the order of $\Lambda_{QCD}$, but the
exact value that should be used is scheme dependent.

The initial
conditions are those of the McLerran-Venugopalan (MV) model~\cite{McLerran:1994ni,*McLerran:1994ka,*McLerran:1994vd};
one has for the initial rapidity $Y=0$, $V(\xt) =
\prod_{k=1}^{N_y}\exp(-i \frac{g\rho_k(\xt)}{\nabla_T^2})$ with
$\langle \rho^a_k(\xt)\rho^b_l(\yt)\rangle = g^2\mu^2 \delta^{(2)}
(\xt-\yt)\delta^{ab}\delta^{kl}/N_y$, with the indices $k,l=1\dots
N_y$ representing a discretized longitudinal coordinate, taking care
of the finite extension of the source in the $x^-$ direction.  The
normalization is chosen such that $\sum_{k,l}\langle
\rho^a_k(\xt)\rho^b_l(\yt)\rangle = g^2\mu^2 \delta^{(2)}
(\xt-\yt)\delta^{ab}$.  Given the numerical implementation of this
initial condition~\cite{Lappi:2007ku} for the $V$'s, using a
longitudinal resolution of $N_y=100$, one can then solve
\eq\nr{eq:Langevin} on a  2-D lattice. The Poisson
equation is solved by leaving out the zero transverse momentum
mode. This procedure corresponds to an infrared cutoff given by the
size of the system.
The calculation is performed on a regular square
lattice of $N_T^2$ sites with periodic boundary conditions. Its volume is $(N_Ta)^2$ where $a$ is the lattice spacing. 
In the calculations presented we use $N_T=512$ unless otherwise stated, and $\delta s=0.00026$, $g^2\mu a=0.109375$.
The parameters controlling the running coupling are taken as
 $c=0.2$, $\Lambda=0.0536\, g^2\mu$, and $\mu_0=2.5 \Lambda$.
These parameters are chosen to be close to the phenomenologically
realistic range for the speed of evolution as observed in 
fits to $F_2(x,Q^2)$ data, but have not been adjusted to give a 
best possible fit.




\section{Naive large $\nc$ approximation}

\begin{figure}[tbh]
\setlength{\unitlength}{3947sp}%
\begin{picture}(1448,1652)(1036,-1930)
\put(1301,-1561){\line(1,0){1000}}
\put(1201,-1461){\line(0,1){1000}}
\put(1201,-361){\circle{150}}
\put(2401,-1561){\circle{150}}
\put(2401,-361){\circle*{150}}
\put(1201,-1561){\circle*{150}}
\put(1801,-1461){\makebox(0,0)[b]{\large$r$}}
\put(1101,-961){\makebox(0,0)[r]{\large$r$}}
\put(1051,-1861){\makebox(0,0)[lb]{\large$\xt$}}
\put(2251,-1861){\makebox(0,0)[lb]{\large$\yt$}}
\put(2251,-61){\makebox(0,0)[lt]{\large$\ut$}}
\put(1051,-61){\makebox(0,0)[lt]{\large$\vt$}}
\end{picture}
\caption{Square arrangement of the coordinates in
  \eq\nr{eq:quad}. The filled circles
represent Wilson lines  and the open ones conjugates.
} \label{fig:sqcoord}
\end{figure}
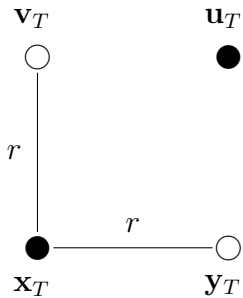

\begin{figure*}[tbh]
\begin{center}
\includegraphics[width=0.47\textwidth]{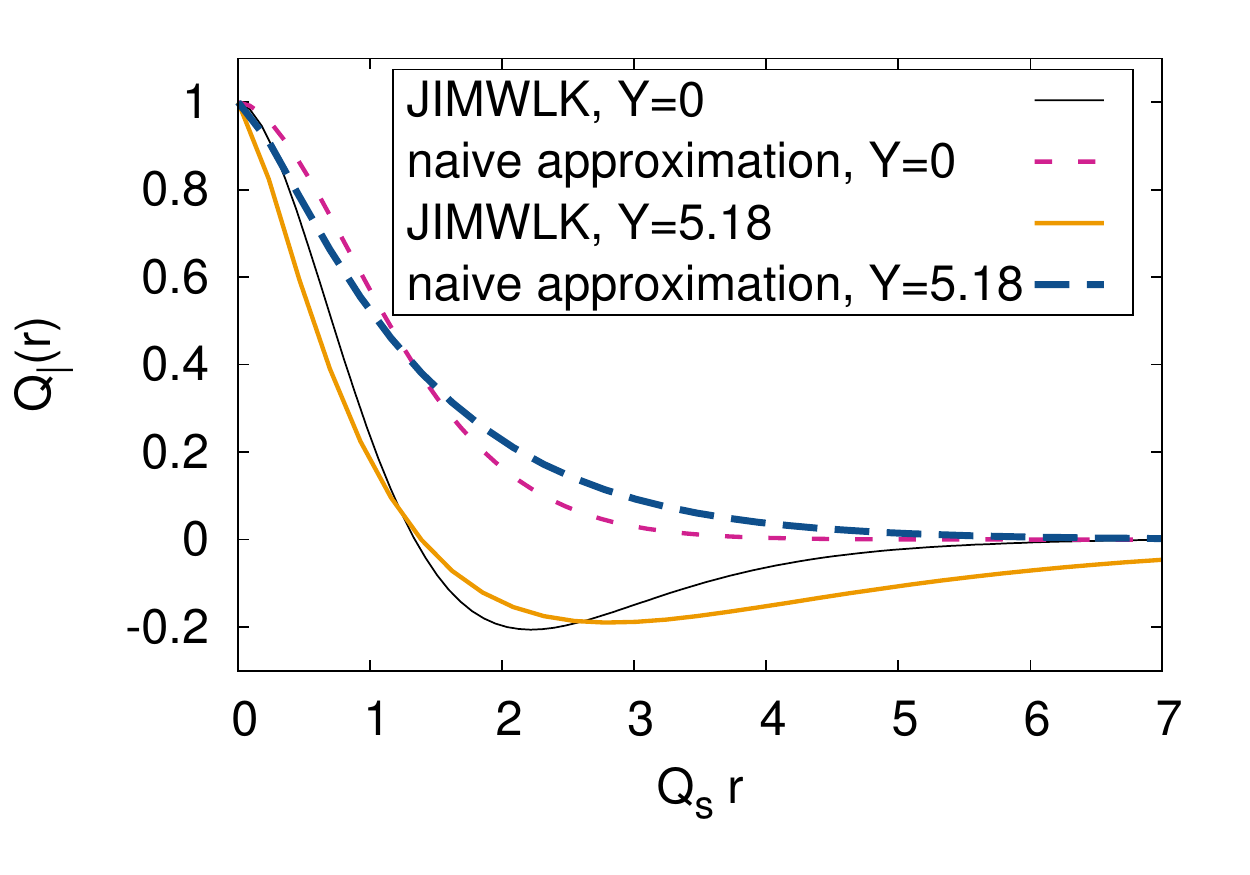}
\includegraphics[width=0.47\textwidth]{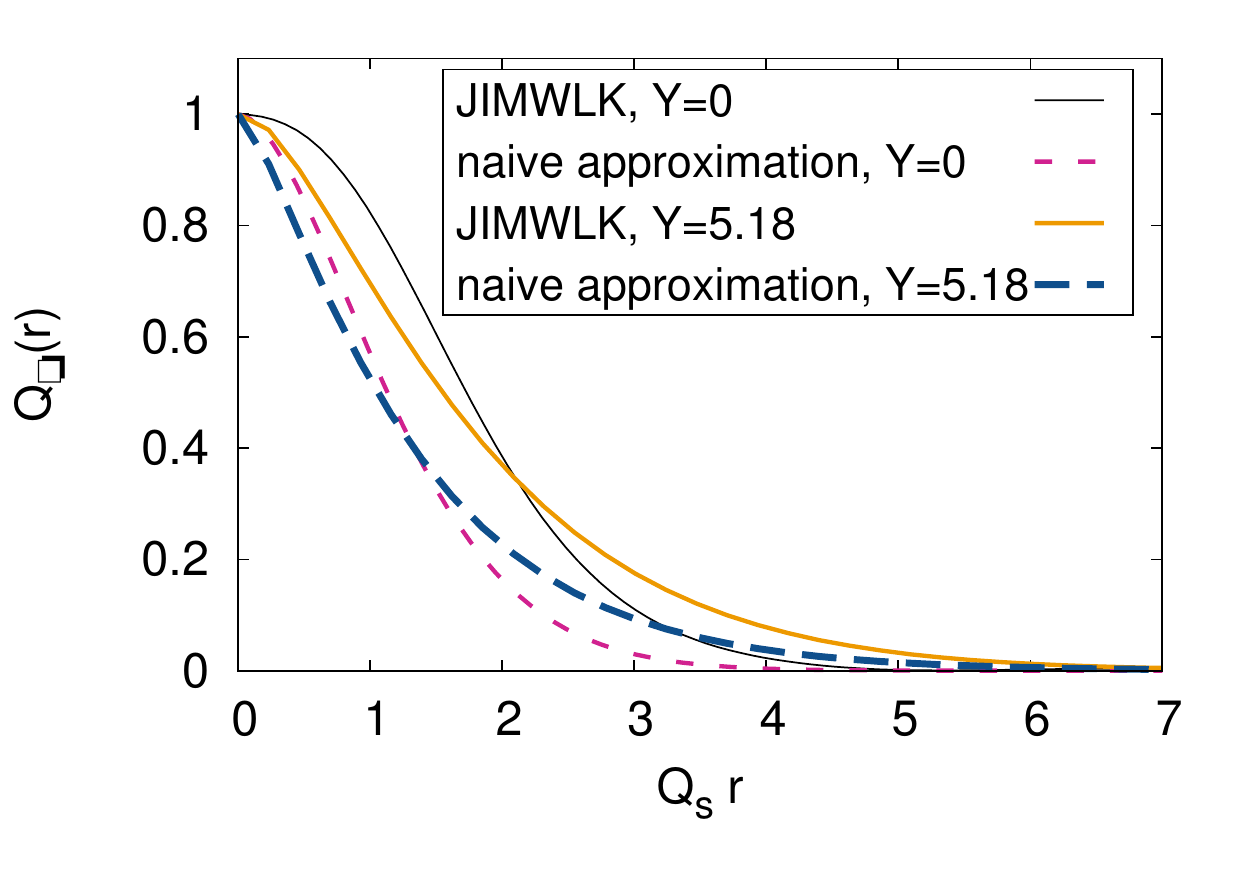}
\end{center}
\vskip -0.3 in
\caption{ The four point function $Q$ for coordinates in the line
  configuration (left) and the square configuration (right) for the
  initial condition ($Y=0$) and after evolution ($Y=5.2$). The
  JIMWLK quadrupole expectation value (solid lines) is compared to the
  naive approximation in \eq\nr{eq:qlargenclinesq} (dashed
  lines). \label{fig:Qnaive} 
}
\end{figure*}

We will first present results for the expectation value of the
quadrupole correlator $Q$. Since this correlator is a function of four
independent two dimensional spatial vectors, we will for simplicity
study its properties for two specific spatial configurations, one of which is
the square configuration $Q_\square(r)$, which has the four
coordinates arranged in a square of size $r=|\xt-\yt|$ as shown in
\fig\ref{fig:sqcoord}. The other is a simple line configuration
$Q_|(r)$, where $\ut=\xt,\vt=\yt$ and $r = |\xt-\yt| = |\ut -\vt|$.
The expectation values of all correlators are computed by averaging 
over all positions on the 2D lattice and 20 different initial configurations.

A ``naive'' large $\nc$ approximation for \eq\nr{eq:quad}
considered previously in phenomenological studies is
\begin{multline}
\label{eq:qlargencgeneral}
Q(\xt,\yt,\ut,\vt) \underset{\nc \to \infty}{\approx} \frac{1}{2}
\big( D(\xt,\yt) D(\ut,\vt)
\\
+  D(\xt,\vt) D(\ut,\yt) \big)
\end{multline}
On inspection, it is apparent that this approximation is problematic
because it does not reduce to all the right ``coincidence limits'';
taking $\ut=\vt$ one has $Q(\xt,\yt,\ut,\ut)= D(\xt,\yt)$, but the
r.h.s. of \eq\nr{eq:qlargencgeneral} reduces to $ \frac{1}{2} \left(
D(\xt,\yt) + D(\xt,\ut) D(\ut,\yt) \right) $ instead. Plots of the
JIMWLK solution for the line and square configurations
respectively compared to the ``naive'' approximation 
\begin{equation}\label{eq:qlargenclinesq}
Q_{|}^{\textrm{naive}}(r) = Q_{\square(r)}^{\textrm{naive}} = D(r)^2
\end{equation}
are shown in
\fig\ref{fig:Qnaive}. It is apparent that the approximation fails
even for the initial condition at $Y=0$, a result that is not
ameliorated by the JIMWLK RG evolution. The disagreement between
the JIMWLK and approximate results is greater for the line configuration
than the square configuration. 

\section{Gaussian approximation}

\begin{figure*}[tbh]
  \begin{center}
    \includegraphics[width=0.47\textwidth]{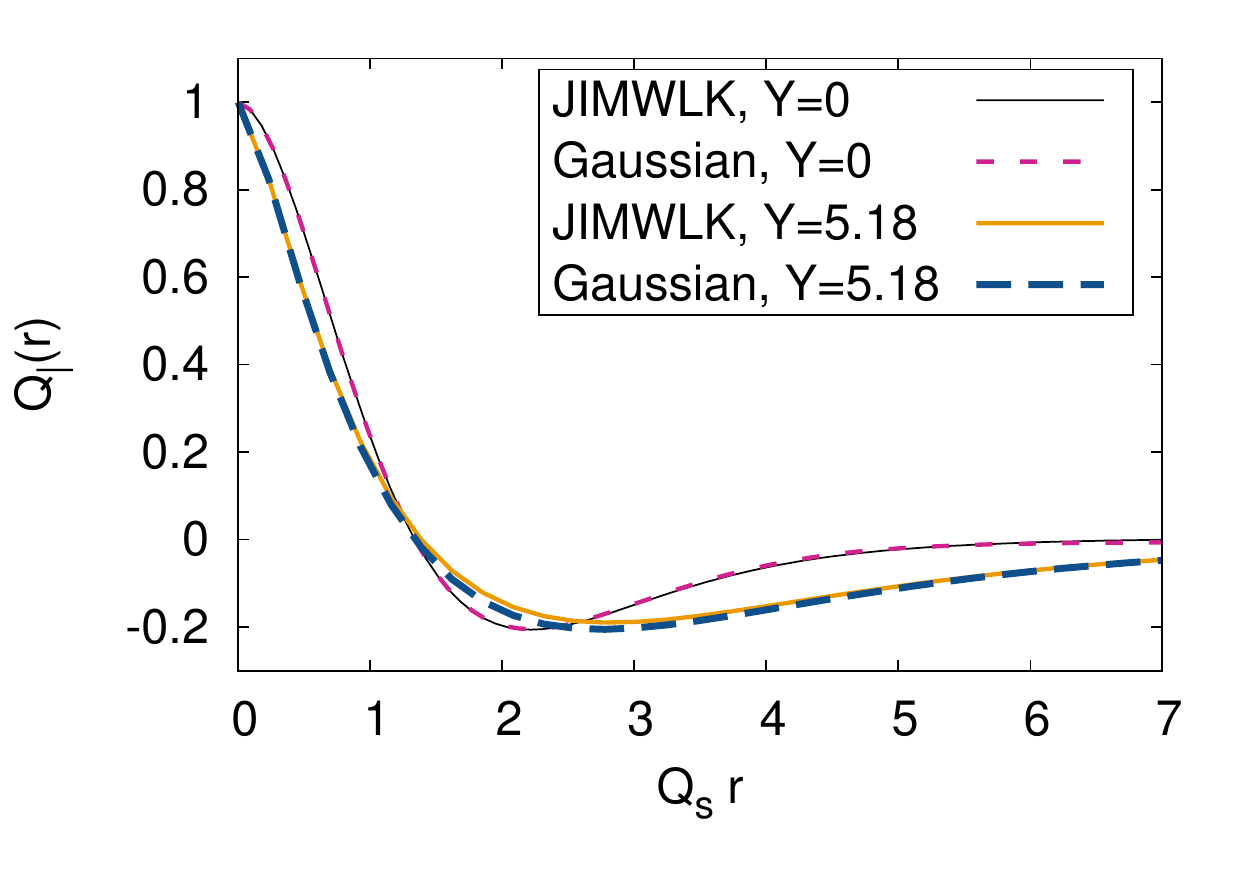}
    \includegraphics[width=0.47\textwidth]{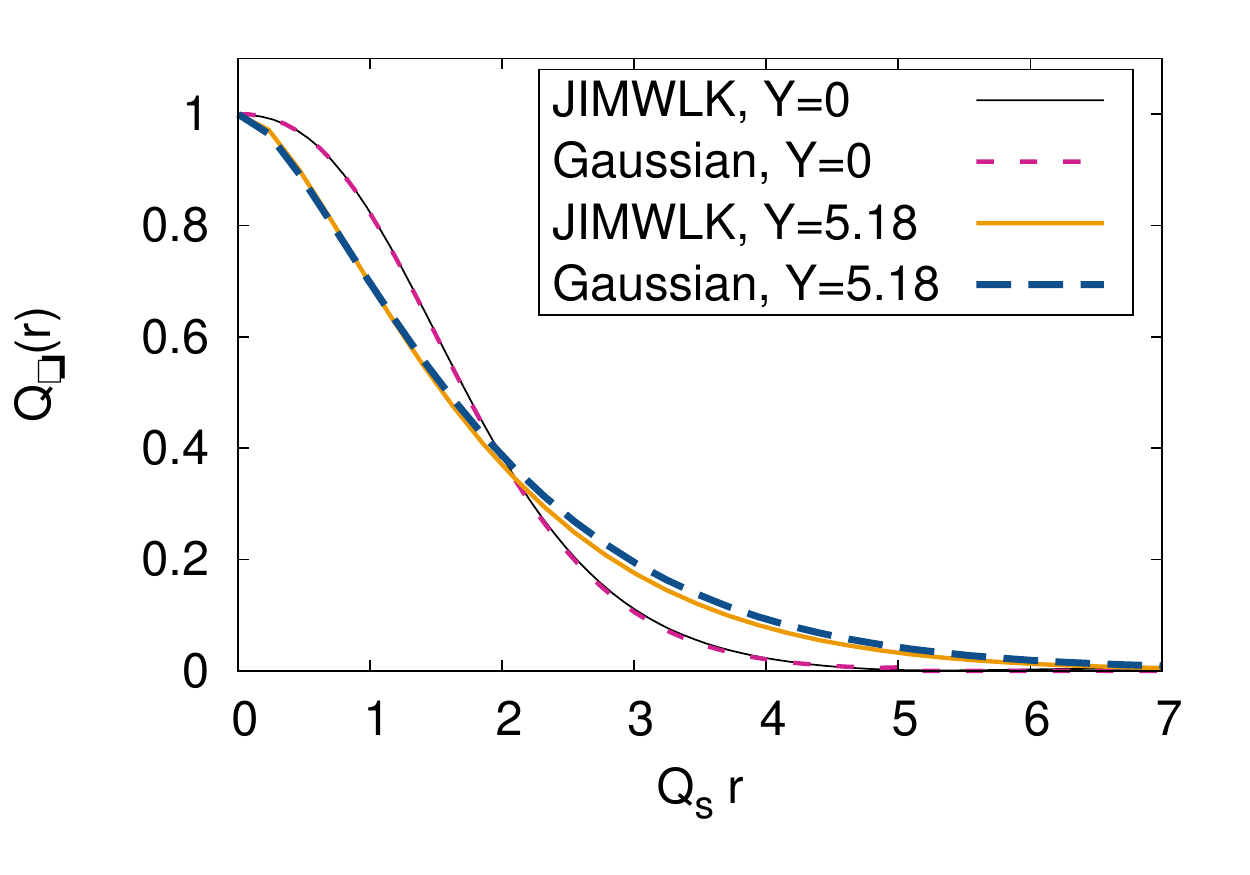}
  \end{center}
  \vskip -0.3 in
  \caption{ The four point function $Q$ for coordinates in the line
    configuration (left) and the square configuration (right) for the
    initial condition ($Y=0$) and after evolution ($Y=5.2$). The
    JIMWLK quadrupole expectation value (solid lines) is compared to
    the Gaussian approximation (dashed lines) Eqs.~\nr{eq:gausq1},~\nr{eq:gausq2}
. \label{fig:QGauss}}
\end{figure*}

The fact that the expectation value of the quadrupole correlator does
not factorize in the naive way of \eq\nr{eq:qlargencgeneral}
 was pointed out in~\cite{Dumitru:2010ak} and 
is seen explicitly already in the MV model that specifies 
the initial conditions\footnote{
We leave the study of quartic contributions~\cite{Dumitru:2011zz} to the MV 
action for future study.}. We will in this paper compare the JIMWLK
result for $Q$ to a ``Gaussian'' 
approximation~\cite{Fujii:2006ab,Marquet:2010cf}
(also referred to as Gaussian 
Truncation~\cite{Weigert:2005us,Kovchegov:2008mk}). This is obtained
by assuming that the correlators of the color charges are Gaussian
variables even after JIMWLK evolution, 
and therefore all the higher point functions can be expressed
in terms of a single two point correlation function,
$ -\ln(D(\xt-\yt)) \equiv  \frac{\cf}{2}\Gamma (\xt-\yt)$
in the notations of ref.~\cite{Dominguez:2011wm}. 
We obtain this two point function from our solution of the 
JIMWLK equation. However, as shown in ref.~\cite{Kovchegov:2008mk},
using the solution of the BK equation for the two point function 
would also be a very good approximation.
The Gaussian approximation  has also been
motivated formally at asymptotically small $x$ in ref.~\cite{Iancu:2002aq}.
Thus in the Gaussian approximation 
the higher point functions are related to the two point function 
similarly as in the MV model. For the specific case of the
quadrupole $Q$ the explicit expression has been derived in 
ref.~\cite{Dominguez:2011wm}.

\begin{figure*}[tbh]
\begin{center}
\includegraphics[width=0.47\textwidth]{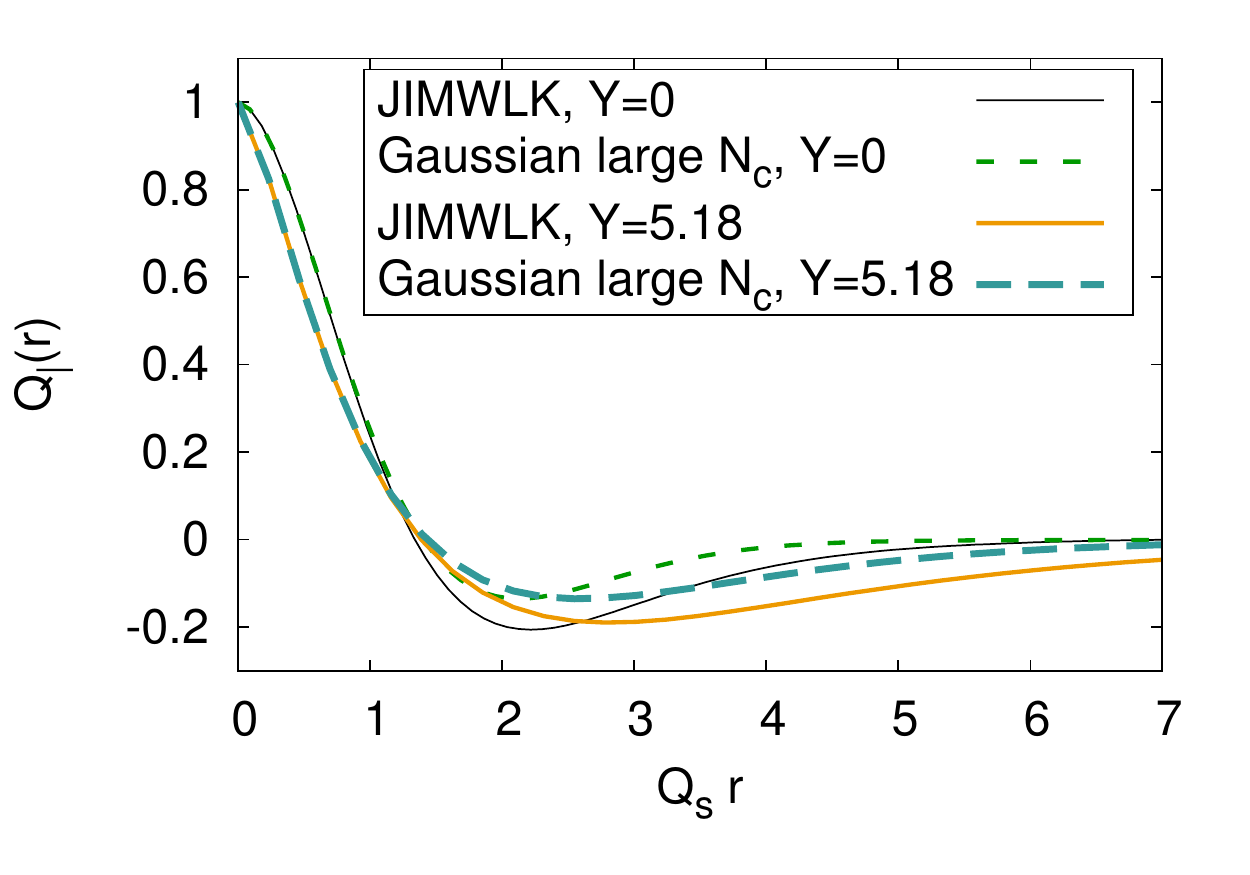}
\includegraphics[width=0.47\textwidth]{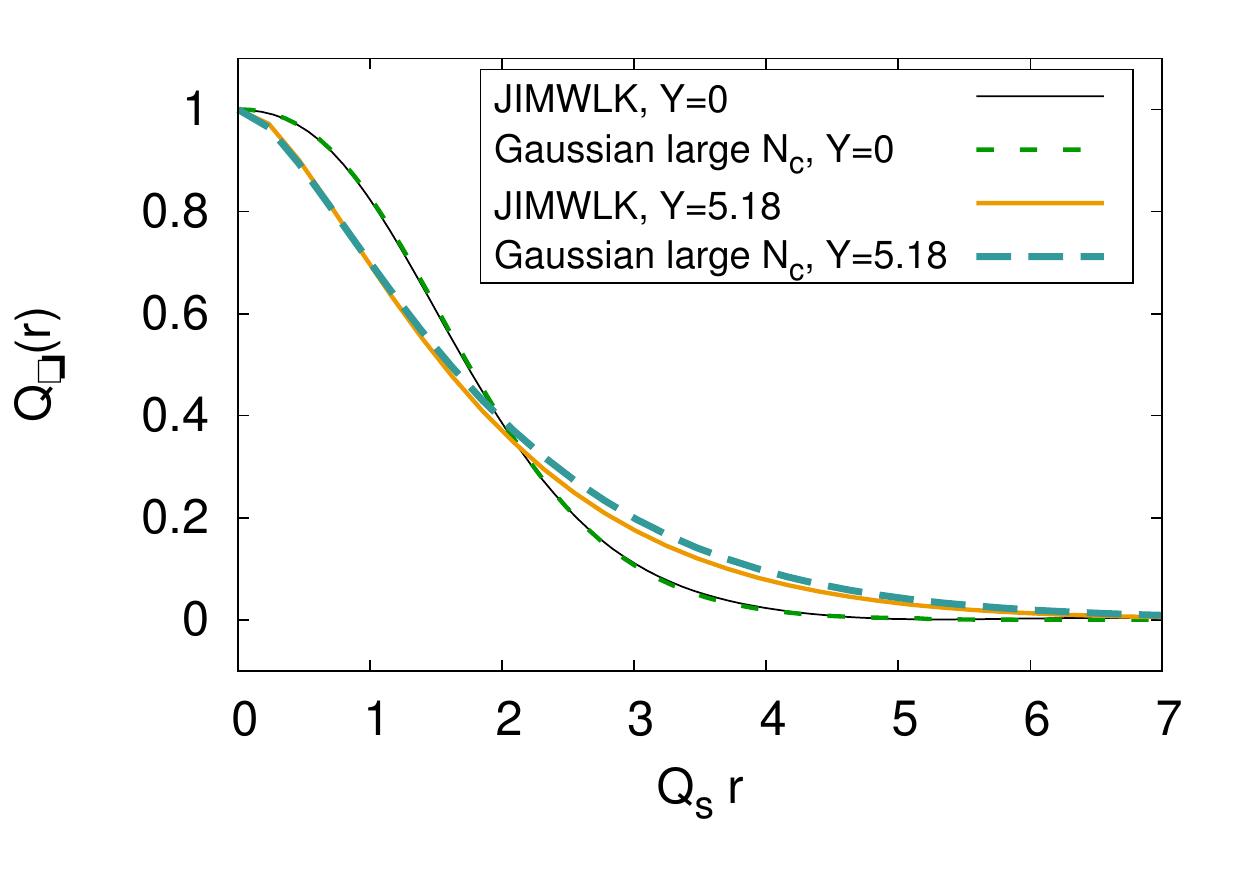}
\end{center}
\vskip -0.3 in
\caption{ The four point function $Q$ for coordinates in the line
  configuration (left) and the square configuration (right) for the
  initial condition ($Y=0$) and after evolution ($Y=5.2$). The
  JIMWLK quadrupole expectation value (solid lines) is compared to the
  large $\nc$ limit of the Gaussian approximation, 
Eqs.~\nr{eq:QdipNcInftyline},~\nr{eq:QdipNcInftysq}
(dashed lines).  }
\label{fig:QGauss-largeN}
\end{figure*}

For the square and line configurations we consider here, 
the cumbersome general Gaussian expression for the quadrupole 
correlator simplifies greatly and we find,
\begin{eqnarray} 
\label{eq:gausq1}
Q_|(r) &\underset{\nc<\infty}{\approx}& \frac{\nc + 1}{2}\Big( D(r)\Big)^{2\frac{\nc+2}{\nc+1}}
\\ \nonumber & &
-\frac{\nc - 1}{2}\Big( D(r)\Big)^{2\frac{\nc-2}{\nc-1}}
\\
\label{eq:gausq2}
Q_\square(r) &\underset{\nc<\infty}{\approx}& \left( D(r)\right)^2 
\bigg[
\frac{\nc + 1}{2}\left( \frac{D(r)}{D(\sqrt{2}r)}\right)^{\frac{2}{\nc+1}}
\\ \nonumber &&
-\frac{\nc - 1}{2}\left( \frac{D(\sqrt{2}r)}{D(r)}\right)^{\frac{2}{\nc-1}}
\bigg].
\end{eqnarray}

In \fig\ref{fig:QGauss}, the numerical results from the solution of
the JIMWLK RG equation for the quadrupole (in the line and square
configurations) are compared to this Gaussian approximation.  When
computing the Gaussian approximation, we chose to only calculate the
averages in $D$ from contributions aligned with the square or the line
configuration, respectively.  While the agreement of $Q$
with its Gaussian approximation for the initial condition $Y=0$
is required by definition (because one has MV initial conditions in
both cases), the agreement of the JIMWLK simulation with the evolved
Gaussian approximation is remarkably good. This suggests that the
computations in ref.~\cite{Fujii:2006ab,Dusling:2009ni} that rely on
this approximation may, at least in this aspect, be robust.

If one takes the large $\nc$ limit of the expressions in
\eqs\nr{eq:gausq1} and~\nr{eq:gausq2}, one finds that
\begin{eqnarray}
 \label{eq:QdipNcInftyline}
Q_|(r) &\underset{\nc\to\infty}{=}& D^2(r) [1 + 2\ln(D(r))]~, \\
 \label{eq:QdipNcInftysq}
Q_\square(r) &\underset{\nc\to\infty}{=}& D^2(r)
\left[1+2\ln\left(\frac{D(r)}{D(\sqrt{2} r)}\right)\right]~.
\end{eqnarray}
The JIMWLK result compared to this particular large $\nc$
approximation is shown in \fig\ref{fig:QGauss-largeN}.  The
agreement of the large $\nc$ Gaussian approximation with the JIMWLK
result is quite good for $Q_\square$, especially for the initial condition; it is less
so for $Q_|$. In the latter case, the agreement for
small $r \qs$ is good, with discrepancies showing up for $r \qs \gtrsim 1$.
 We note that the difference between the ``naive'' 
approximation~\nr{eq:qlargenclinesq} and
this Gaussian large $\nc$ approximation \nr{eq:gausq1},~\nr{eq:gausq2}
is in the additional
logarithmic term in the latter. In the case of the line configuration,
this provides some insight into the discrepancy of the naive
approximation with the result from JIMWLK evolution. The latter is not constrained to
be greater than zero, while the former is. The additional logarithmic
term relaxes this constraint because it can change the sign of the
result.

\begin{figure*}[tbh]
\begin{center}
\includegraphics[width=0.47\textwidth]{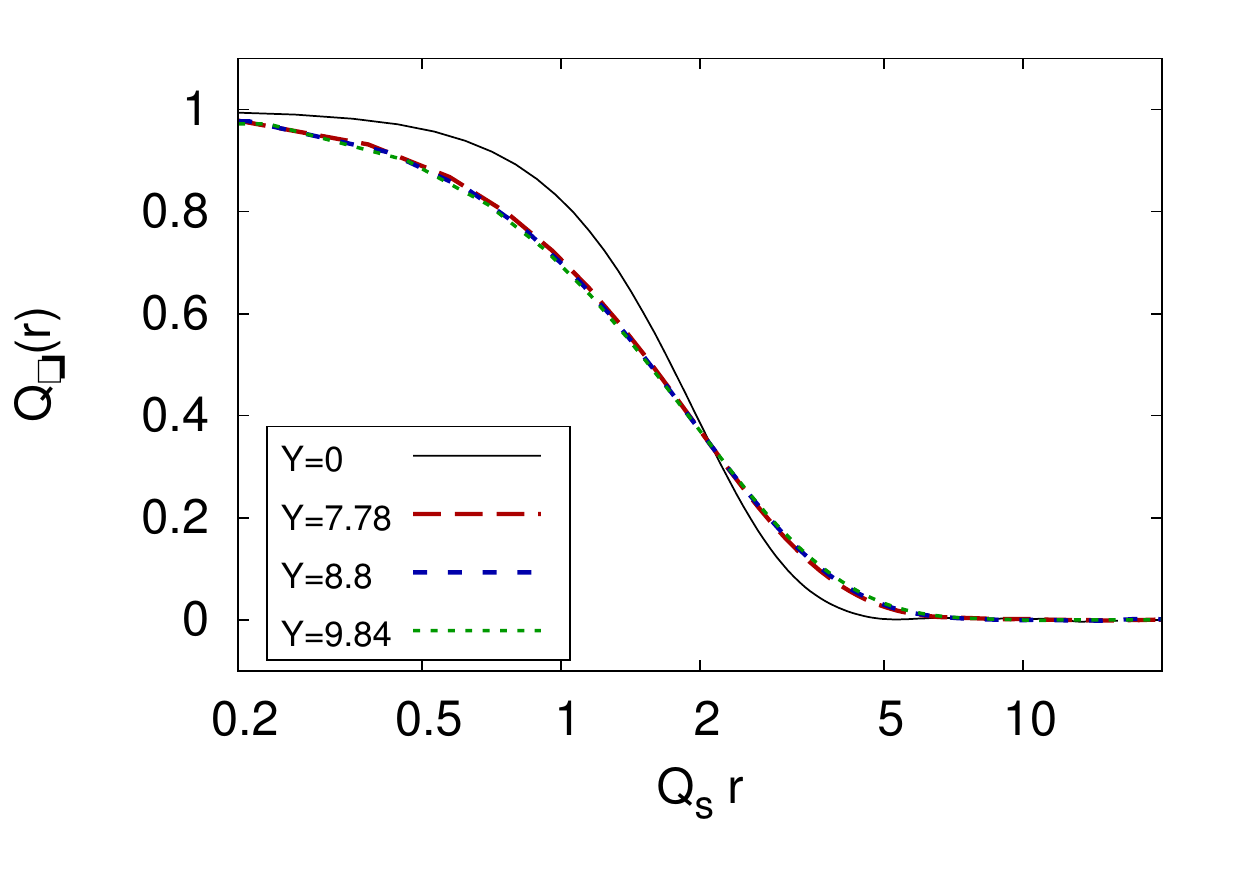}
\includegraphics[width=0.47\textwidth]{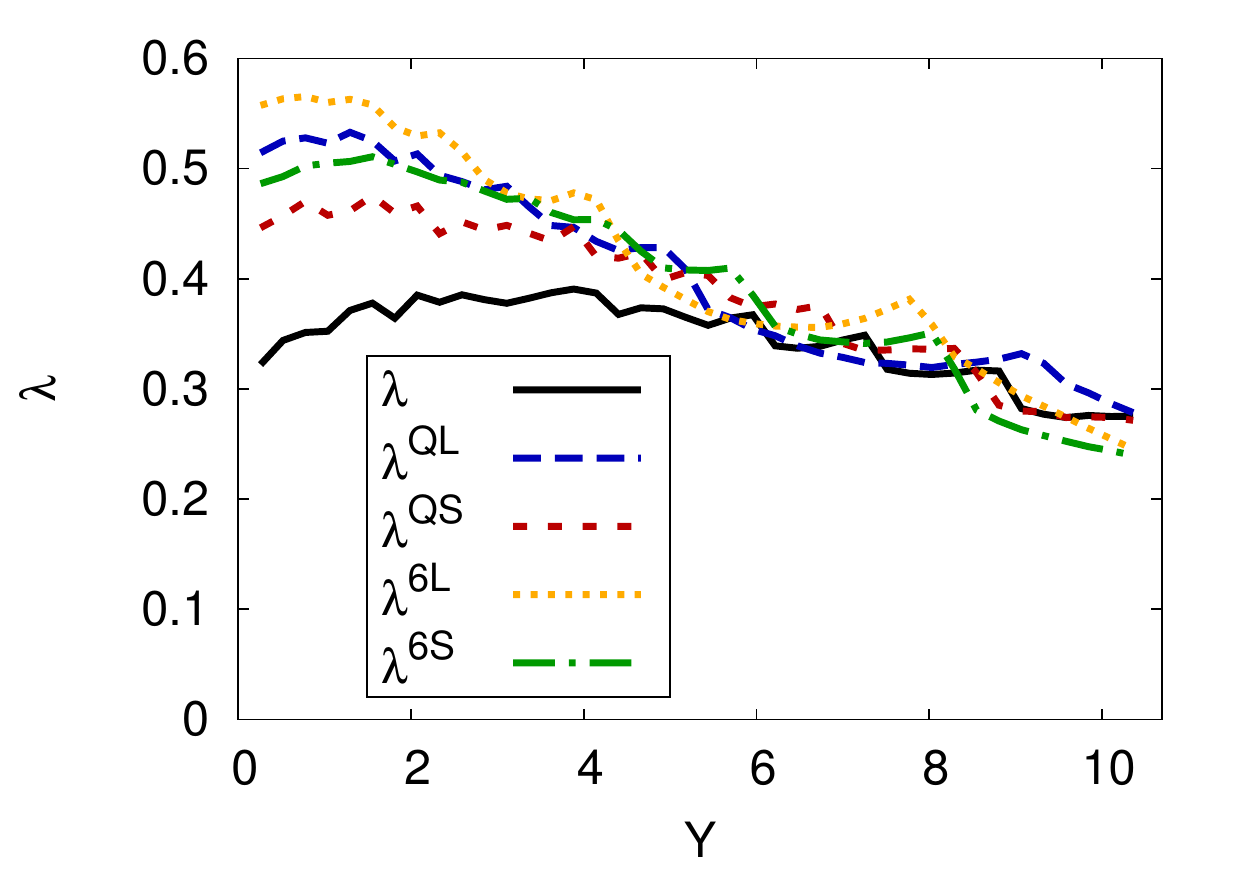}
\end{center}
\vskip -0.3 in
\caption{ Left: The JIMWLK quadrupole amplitude $Q_\square$
  for the square configuration versus the scaling variable $\qs r$
  (on a logarithmic axis).
  After initial transient behavior, the amplitude settles to
  a universal curve (for $Y \geq 7.8$) which depends on $\qs r$
  alone. Right: The evolution speed $\lambda = \ud\ln \qs^2/\ud Y$ extracted
  from the dipole amplitude $D$, the quadrupole
  amplitude in the two spatial configurations (line and square, 
  $\lambda^{\mathrm{QL}},\lambda^{\mathrm{QS}}$)
  and the six point function $S_6$, also in the
  two spatial configurations ($\lambda^{6L},\lambda^{6S}$).
 }
\label{fig:geom-scale}
\end{figure*}

\section{Geometric scaling of the quadrupole}

We now turn from this comparison of the quadrupole expectation value
to study aspects of its evolution determined from the solution of the
JIMWLK equation. The RG evolution of the quadrupole has been studied
previously~\cite{JalilianMarian:2004da,Dominguez:2011gc} in the large
$\nc$ limit of Mueller's dipole model~\cite{Mueller:1993rr}. In
particular, it has been argued very recently in
ref.~\cite{Dominguez:2011gc} that quadrupole evolution should
demonstrate geometrical scaling~\cite{Stasto:2000er,*Iancu:2002tr,*Mueller:2002zm}.

\begin{figure*}[htb]
\begin{center}
\includegraphics[width=0.47\textwidth]{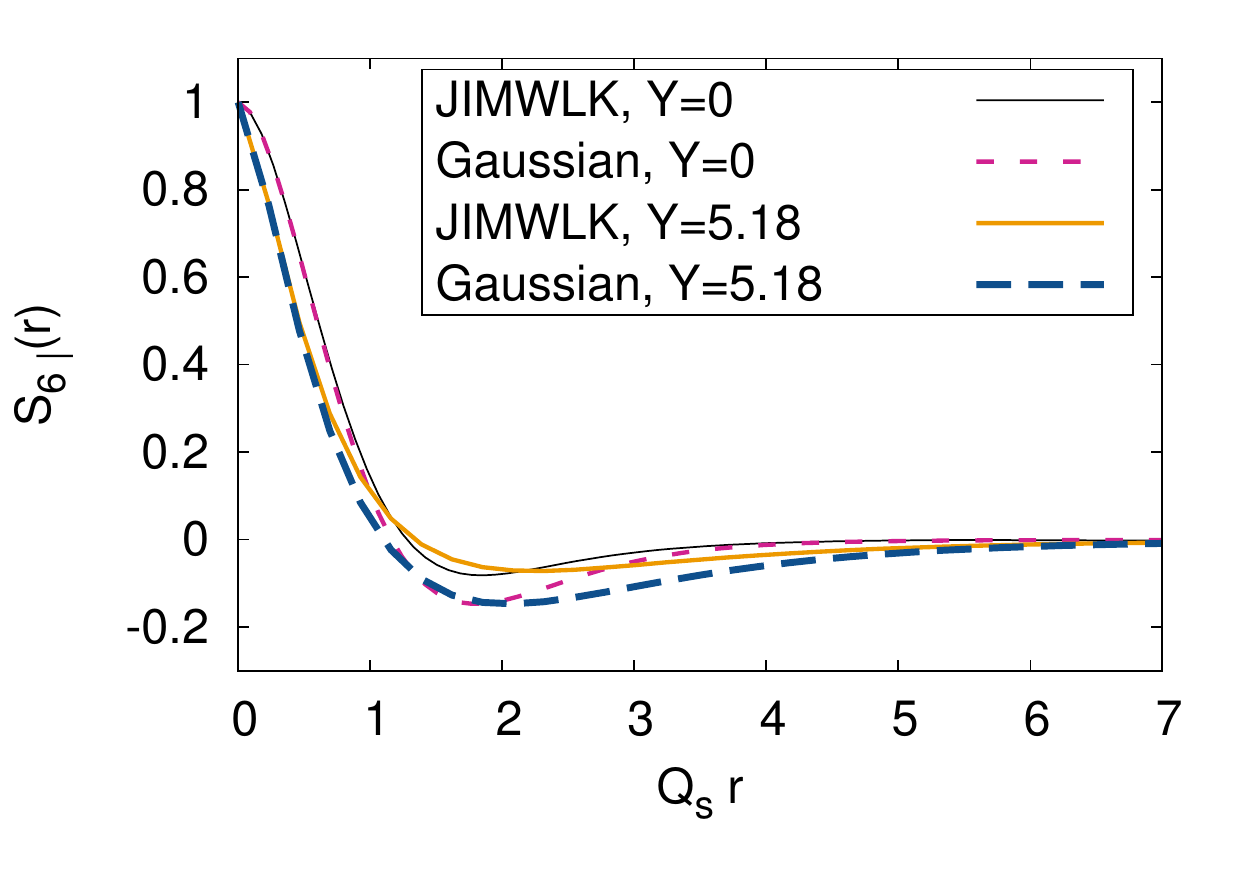}
\includegraphics[width=0.47\textwidth]{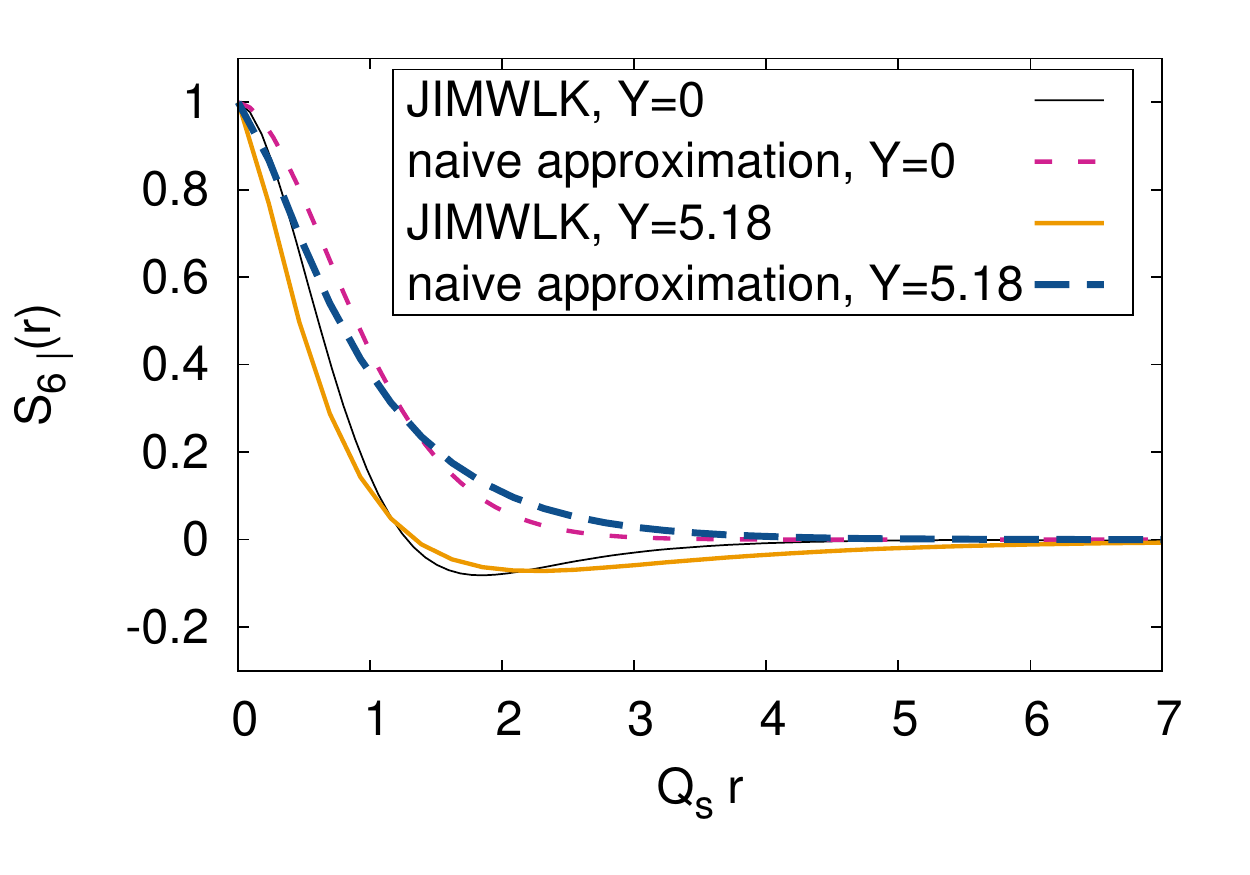}
\end{center}
\vskip -0.3 in
\caption{Left: The JIMWLK result for $S_6$ (see text) plotted for the
  line configuration compared to the Gaussian approximation for
  $S_6$. Right: The JIMWLK result compared to a naive large $\nc$
  result.}
\label{fig:sixpoint-Gauss}
\end{figure*}

Our results are shown in \fig\ref{fig:geom-scale}.
On the left  we plot $Q_\Box$ as a function of $r \qs$.
We observe that after initial transient behavior the amplitude
$Q_\square$ settles on a shape that is a universal function of $r \qs$.
Thus for a given $x$ and $Q^2$ probed in a process, the quadrupole
amplitude depends only on the combination proportional to $\qs(x)^2/Q^2$
thereby demonstrating geometrical scaling for this quantity. 

We previously defined the usual saturation scale $\qs$ through the dipole
operator as $D\left(r=\sqrt{2}/\qs\right) = e^{-1/2}$.
One can analogously characterize the evolution 
of the quadrupole by introducing ``quadrupole saturation scales''
corresponding to the square and line coordinate arrangements. We define
these as 
$Q_{|,\Box}\left(r=\sqrt{2}/\qs^{|,\Box}\right) = e^{-1/2}$.
On the right of \fig\ref{fig:geom-scale} we show the evolution speeds
$\lambda = \ud\ln \qs^2/\ud Y$ of these different saturation scales,
$\lambda$ for the dipole and $\lambda^{QL},\lambda^{QS}$ for
the quadrupole in the line and square configurations respectively.
 The plot shows that initially the quadrupole evolves more
rapidly than the dipole. 
After about 6 units in rapidity, the evolution reaches
a universal geometrical scaling regime and  the $\lambda$
parameters settle to a common value. 

\section{Six point function and dihadron correlations}

Let us now return to the expression we had in \eq\nr{eq:pA} for forward
di-hadron production in hadronic collisions. Experiments at RHIC for
deuteron-gold scattering at high energies have shown that the
away-side peak in di-hadron correlations is significantly broadened
for central collisions at forward
rapidities~\cite{Adare:2011sc,*Braidot:2011zj} as predicted in the CGC
framework~\cite{Marquet:2007vb} and confirmed in more detailed
analyses~\cite{Albacete:2010pg,*Tuchin:2009ve}. However, these analyses
relied on factorization assumptions that, as we have seen, are not
justified because the quadrupole correlator is not simply
factorizable. It is not the quadrupole correlator that appears
directly in \eq\nr{eq:pA}, so we shall now consider the expression
\begin{multline}
S_6(\xt, \yt,\ut,\vt ) = \frac{\nc^2}{\nc^2-1}\\
\times
\bigg\langle 
\hat{Q}(\xt,\yt,\ut,\vt)
 \hat{D}(\vt,\ut) 
-\frac{1}{\nc^2}\hat{D}(\xt,\yt)
 \bigg\rangle,
\end{multline}
 corresponding to
the first and last terms in the brackets in \eq\nr{eq:pA}. The
JIMWLK evolution equation for this quantity was derived
recently~\cite{Dumitru:2010ak}. As for the quadrupole correlator, we
would like to compare numerical results for $S_6$ to a Gaussian
approximation. The latter, however, would strictly require that one
compute the product $\langle \hat{Q} \hat{D}\rangle$ in this approximation. 
Since
results for this quantity are not at present available, we will assume
that $\langle \hat{Q} \hat{D}\rangle\approx 
\langle \hat{Q}\rangle \langle \hat{D}\rangle$ and
compare the $\nc=3$ Gaussian approximation for 
$\langle \hat{Q}\rangle \langle \hat{D}\rangle$ 
to the numerical JIMWLK results for
$S_6$. Figure~\ref{fig:sixpoint-Gauss} (left) shows the result for the
line configuration for $S_6$.  We observe that the agreement of the
JIMWLK and approximate results is quite good for $r \qs \ll1 $ or $r \qs
\gg 1$, but that there are noticeable deviations in the region 
$1 \lesssim r\qs \lesssim 3$.  
Due to the good agreement of $Q$ with the
Gaussian approximation from Eqs.~\nr{eq:gausq1},~\nr{eq:gausq2}
shown above, we interpret these deviations for $S_6$ as ${\cal
  O}(1/\nc)$ corrections to $\langle \hat{Q} \hat{D}\rangle
\approx \langle \hat{Q}\rangle \langle \hat{D}\rangle$. 
For the square configuration of $S_6$ (not shown)
the deviations are much less, as was the case for the quadrupole.
In \fig\ref{fig:sixpoint-Gauss}
(right), we plot the JIMWLK results against the ``naive'' large $\nc$
approximation $S_{6|}(r) \approx S_{6\square}(r) \approx D(r)^3 $
that has previously been considered in the
literature. Once again the deviations are large, suggesting that this
approximation is not tenable.
We characterize the six-point function by the
saturation scales $\qs^{6|,\square},$ defined 
by $S_{6|,\square}(r=\sqrt{2}/\qs^{6|,\square})= e^{-1/2}$. The 
corresponding evolution speeds are also shown in 
\fig\ref{fig:geom-scale}.

As a final result, we present a visualization of JIMWLK evolution that
demonstrates the role of fluctuations. In high energy QCD,
fluctuations from event-to-event can occur because of fluctuations in
the impact parameter positions of gluons, in their position in
rapidity, and in the fluctuations in the number of
gluons~\cite{Miettinen:1978jb}. All of these fluctuations are captured
in the numerical simulations of the JIMWLK hierarchy. These results
are shown in \fig\ref{fig:anim}, which shows the fluctuations of the 
Wilson lines in the transverse plane at different rapidities.
The decreasing of the correlation length $\sim 1/\qs$ with energy 
is clearly visible.
  \begin{figure*}
  \hspace{-1.2cm}
    \begin{minipage}[t]{0.33\linewidth}
      \centering
      \animategraphics[loop,autoplay,width=7cm,controls,trim=1cm 3.5cm 0cm 2cm]{5}{./anim/d}{000}{050}\\\vspace{0.4cm}
    \end{minipage}
    \hspace{-0.8cm}
   \begin{minipage}[t]{0.33\linewidth}
      \centering
      \animategraphics[width=7cm,trim=1cm 3cm 0cm 2cm]{6}{./anim/d}{025}{025}\\\vspace{0.2cm}
    \end{minipage}
     \hspace{-0.8cm}
    \begin{minipage}[t]{0.33\linewidth}
      \centering
      \animategraphics[width=7cm,trim=1cm 3cm 0cm 2cm]{6}{./anim/d}{050}{050}\\\vspace{-0.2cm}
    \end{minipage}
    \caption{(animated online, requires Acrobat reader) Correlation $1/\nc\langle V^\dag(0,0) V(x,y)\rangle$ between the center position $(0,0)$ and the point $(x,y)$ for three different rapidities $Y$. This illustrates the degree of fluctuations and shows how the correlation length
      decreases dynamically with increasing $Y$. The first image can be animated to show the evolution with rapidity.\label{fig:anim}}
   \end{figure*}

To summarize, we have in this work performed simulations of the
running coupling SU(3) JIMWLK equation that describes the behavior of
expectation values of Wilson line correlators in high energy QCD. We
have presented first results for the evolution of specific higher
$n$-point functions which are related to experimental observables in
DIS and in p+p, p+A collisions. In particular, we find that the
``quadrupole'' correlator can be approximated quite well by a careful
Gaussian approximation for $\nc=3$. The $\nc\to\infty$ Gaussian
approximation is accurate at short distances, $r\qs \lesssim 1$, but may
display more significant relative deviations in the saturation regime,
$r\qs\gtrsim 1$. We have also provided evidence for travelling wave
solutions and geometric scaling for the quadrupole.

\section*{Acknowledgements}
Discussions with G.\ Beuf, F.\ Dominguez, F.\ Gelis, E.\ Iancu, A.~H.\ Mueller,
K.\ Rummukainen, H.\ Weigert, D.\ Triantofyllopoulos, B.-W.\ Xiao and
F.\ Yuan are gratefully acknowledged. T.L.\  has been supported by the
Academy of Finland, projects 126604 and 141555 and by computing
resources from CSC -- IT Center for Science in Espoo, Finland. He
thanks the Nuclear Theory Group at BNL for its hospitality during the
early stages of this work. B.P.S.\ and R.V.\ are supported by US
Department of Energy under DOE Contract No.DE-AC02-98CH10886.
A.D.\ and J.J.-M.\
acknowledge support by the DOE Office of Nuclear Physics
through Grant No.\ DE-FG02-09ER41620 and from The City University of New York through the PSC-CUNY Research
Award Program, grants 63382-0041 (A.D.) and 63404-0041 (J.J-M.).
We acknowledge support from the ``Lab Directed Research
and Development'' grant LDRD~10-043 (Brookhaven National Laboratory),

\bibliography{spires}
\bibliographystyle{JHEP-2modM}

\end{document}